\documentclass[aps,prb,superscriptaddress,showpacs,floatfix,twocolumn]{revtex4-1}
\usepackage{times}
\usepackage{graphicx,graphics,color}
\usepackage{amsmath}
\usepackage{amssymb}
\usepackage{ulem} 
\usepackage{color}

\newcommand{\FeSe}{$A_x$Fe$_{2-y}$Se$_2$}

\begin{document}

\title{Fermi surface evolution and checker-board block-spin antiferromagnetism in $A_x$Fe$_{2-y}$Se$_{2}$}

\author{Yuan-Yen Tai}
\affiliation{Department of Physics, University of Houston, Houston, Texas 77004, USA}
 
\author{Jian-Xin Zhu}
\affiliation{Theoretical Division, Los Alamos National Laboratory, Los Alamos, New Mexico 87545, USA}

\author{Matthias J. Graf}
\affiliation{Theoretical Division, Los Alamos National Laboratory, Los Alamos, New Mexico 87545, USA}

\author{C. S. Ting}
\affiliation{Department of Physics, University of Houston, Houston, Texas 77004, USA}

\date{\today}

\begin{abstract}
We develop an effective multiorbital mean-field  $t$-$J$ Hamiltonian with realistic tight-binding and exchange parameters to describe the electronic and magnetic structures of iron-selenide based superconductors $A_x$Fe$_{2-y}$Se$_2$  for iron vacancy doping in the range $0 \leq y \leq 0.4$.
The Fermi surface topology extracted from the spectral function of angle-resolved photoemission spectroscopy (ARPES) experiments is adequately accounted for by a tight-binding lattice model with random vacancy disorder.
Since introducing iron vacancies breaks the  lattice periodicity of the stochiometric compound, it greatly affects the electronic band structure.  With changing vacancy concentration the electronic band structure evolves, leading to a reconstruction of the Fermi surface topology.
For intermediate doping levels, the realized stable electronic structure is a compromise between
the solutions for the perfect lattice with $y=0$ and the vacancy stripe-ordered lattice with $y=0.4$, which results in a competition between vacancy random disorder and vacancy stripe order. 
A multiorbital hopping model is parameterized by fitting Fermi surface topologies to ARPES experiments,
from which we construct a mean-field $t$-$J$ lattice model to study the paramagnetic and antiferromagnetic (AFM) phases of K$_{0.8}$Fe$_{1.6}$Se$_2$.
In the AFM phase the calculated spin magnetization of the $t$-$J$ model leads to a checker-board block-spin structure in good agreement with neutron scattering experiments and {\it ab}-{\it initio} calculations.
\end{abstract}

\maketitle

\section{Introduction} 

The discovery of high-$T_c$ superconductivity in iron-selenide superconductors \FeSe\  ($A=$K, Rb, Cs, Tl, Tl/K, Tl/Rb)~\cite{JGuo:2010,AKMaziopa:2011,MFang:2011,HWang:2011} has generated increased excitement in the search for high-temperature superconductivity. It provides a new opportunity to understand the underlying physics of iron-based superconductors. The new family of compounds has several unique features:
(i) Superconductivity (for $x \sim 1$ and $y \sim 0.12-0.3$) emerges in proximity to an insulating phase~\cite{MFang:2011,DMWang:2011} (for $x \sim 0.8$ and $y \geq 0.4$), instead of a poor metal as in other iron-based parent compounds.  For the iron deficient compounds with $y\geq 0.4$, there is mounting evidence for the existence of iron vacancy ordered superstructures stabilized with a stripe-like antiferromagetic (AFM) state.~\cite{LHaggstrom:1991,ZWang:2011,XWYan:2011,CCao:2011}  
This raises the interest in the possibility of the insulating phase being driven by the Mott localization~\cite{RYu:2011,YZhou:2011} due to the reduction in kinetic energy.~\cite{CCao:2011}  In particular, the compounds with $x\sim 0.8$ and $y \sim 0.4$ are of special interest for the formation of a peculiar vacancy order (so called $\sqrt{5} \times \sqrt{5}$ superstructure)
as well as a block-spin antiferromagnetic (BAFM) state.\cite{WBao:2011,CCao:2011b,FYe2011,Pomjakushin2011}  
(ii) The end members of the series AFe$_2$Se$_2$ ($x=1$ and  $y=0$)
are heavily electron doped (0.5 electron/Fe) relative to other iron-based superconductors (such as LaOFeAs, BaFe$_2$Se$_2$, FeSe etc.). Band structure calculations~\cite{LZhang:2009,XWYan:2011b,CCao:2011c,IShein:2011,INekrasov:2011}
for these end compounds show only electron pockets that are primarily located
 around the $M$ point of the Brillouin zone (BZ) as defined for a simple tetragonal structure.
Indeed a series of angle-resolved photoemission spectroscopy (ARPES) experiments
has been performed on recently discovered superconducting compounds 
$A_x$Fe$_{2-y}$Se$_2$.~\cite{YZhang01,HDing01} 
A common feature is the presence of electron pockets  around the $M$ point in the Brillouin zone and  a marked absence or near absence of a hole pocket at the $\Gamma$ point.
With regard to electron pockets, the FeSe-122 family is similar to the isostructural FeAs-122 family, while they differ with respect to the hole pocket, which is considered to be essential for certain interband pairing models of superconductivity.

These unique features raise the hope to gain new insights into the mechanism of iron-based superconductivity by studying the FeSe-122 family.
However, special care must be taken in the interpretation of these results due to the complicated real-space structure of highly iron deficient compounds.
The real-space structure for different  Fe compositions  ($0 \le y \le 0.4$) is quite intricate and resembles more that of an alloy than of a lightly doped crystal.
For example, although both  $A$Fe$_2$Se$_2$ ($y=0$) and $A_{0.8}$Fe$_{1.6}$Se$_2$ ($y=0.4$) compounds have perfect lattice periodicity 
(albeit the latter one exhibits striped vacancy order)
in the iron layer,  the lattice structure for compounds with $0 \le y \le 0.4$ 
can be thought of as a superposition of both lattices, which forms either a
random-vacancy lattice or phase-separated lattice with vacancy stripe order.
More generally,  the serious problem of iron deficiency is that the introduced random-disorder scattering centers in the iron layer destroy the translational periodicity, thus rendering the wave vectors of the Bloch wave functions as `bad' quantum numbers to describe electron motion.
Therefore, when interpreting the electronic structure as measured by ARPES, which probes the momentum space,  a real-space electronic structure approach must be developed to account for the strong disorder.

In this paper, we present a systematic study of the evolution of the normal-state electronic structure
and magnetic properties  with vacancy doping based on a real-space description. 
The Fermi surface topology extracted from the spectral function of the ARPES experiments is adequately accounted for by a tight-binding lattice model with random vacancy order at the Fe sites.
We find that introducing Fe vacancies breaks the lattice periodicity and drastically affects the electronic band structure and Fermi surface topology.  The evolution in the band dispersion results in a noticeable reconstruction of the Fermi surface (the technical detail is provided in the Appendix).
As a consequence, for intermediate iron vacancy concentrations,  the realized stable electronic structure is a compromise between
the solutions for $y=0$ (perfect lattice) and $y=0.4$ (stripe ordered lattice), resulting in a competition between vacancy random disorder and vacancy stripe order. Based on such a parameterized hopping model,  the constructed mean-field $t$-$J$ lattice model gives rise to 
a checker-board block-spin structure for K$_{0.8}$Fe$_{1.6}$Se$_2$, which is in good agreement with neutron scattering experiments and {\it ab}-{\it initio} calculations.

The outline of this paper is as follows. 
In Sec.~\ref{SEC:Hamil} we formulate a tight-binding $t$-$J$ model Hamiltonian and introduce within the mean-field approach the Bogoliubov-de Gennes (BdG) equations. 
In Sec.~\ref{SEC:Bloch} we discuss the Bloch wave function formulation of multiorbital electron hopping to obtain a single set of model parameters for the kinetic energy part of the Hamiltonian by fitting the electronic structure and Fermi surfaces of K$_x$Fe$_2$Se$_2$ with a perfect lattice structure.
In Sec.~\ref{SEC:random} we discuss the electronic structure of a random vacancy lattice by introducing an auxiliary impurity scattering approach in the unitarity limit.  In the case of the supercell calculations, the results are in good agreement with the Bloch wave function method. 
In Sec.~\ref{SEC:AFM} we present our calculations of the magnetic structure, which agree well with the neutron scattering measurements. 
The summary is given in Sec.~\ref{SEC:Conclusion}.

\section{Model and  Formalism}
\label{SEC:Hamil}

\begin{figure}  
 \includegraphics[scale=0.45,angle=0]{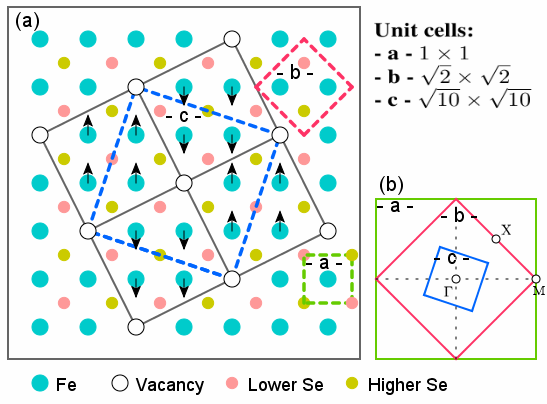}
  \caption
   {(Color online) Different cuts (unit cells) of the real-space lattice of FeSe-122 (a) and corresponding unit cells (same color map) in the Brillouin zone (b).
   The arrows indicate the block-spin AFM structure. For unit cells with cuts ``-b-'' and ``-c-'' the Se atoms in the layers above and below are shown. 
   In the vacancy-ordered state ($y=0.4$) vacancies phase-separate to form stripes. } 
   \label{FIG:structure}
\end{figure}

In this work, we adopt the tight-binding model Hamiltonian that successfully describes the structurally related FeAs-122 superconductors.
Zhang and coworkers \cite{DZhang01} suggested that the mediation of upper and lower As atoms could lead to different hopping terms between iron atoms in the iron layer.
It has been noted that this Hamiltonian has a lower S$_{4}$ point group symmetry with respect to the As atoms as compared to the tetragonal $C_{4}$ symmetry of the bulk crystal, which should give rise to additional exotic magnetic and superconducting phases.\cite{JHu2012}
Since this tight-binding model was introduced, several successful studies have been performed \cite{TZhou} to 
describe  ARPES,\cite{KTerashima09,YSekiba09}
magnetic structures,\cite{Cruz09, TMChuang10}  phase diagrams,\cite{DKPratt09} and vortex core and spin susceptibility in RPA calculations.~\cite{YGao01,YGao02}

\begin{figure}
 \includegraphics[scale=0.35,angle=0]{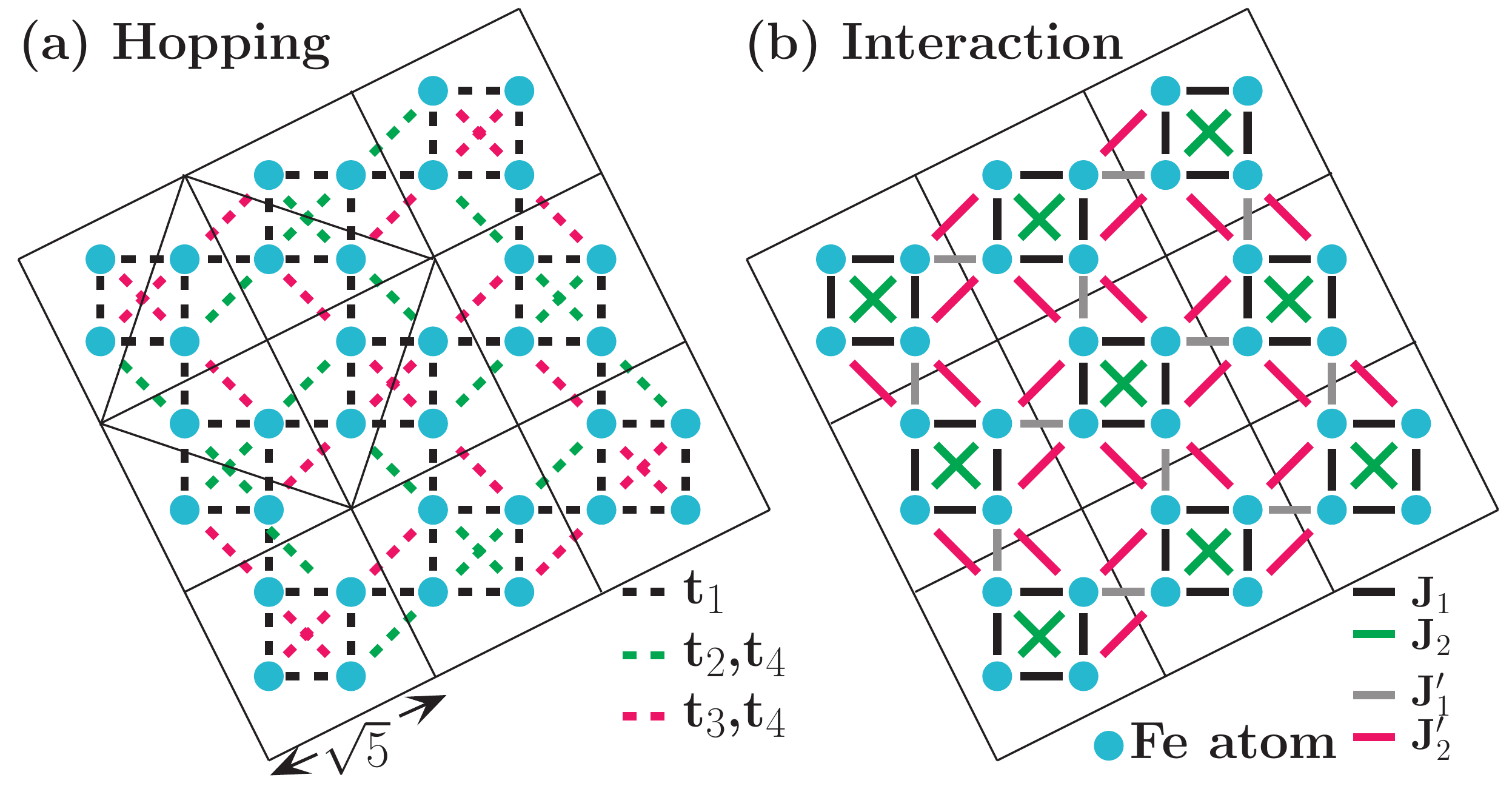}
  \caption
   {(Color online) Schematic pictures of the effective $t$-$J$ Hamiltonian.
   (a) The effective tight-binding model, where $t_1$ is the nearest-neighbor ($nn$) hopping,
   $t_2$ ($t_3$) are the next-nearest-neighbor ($nnn$) intra-orbital hopping terms due to up (down)
   Se atoms and $t_4$ is the $nnn$ interorbital hopping.
   (b) The effective exchange interactions, where
   $J_1$ and $J_2$ are the $nn$ and $nnn$ interactions inside each block, while
   $J_1'$ and $J_2'$ are the $nn$ and $nnn$ interactions between blocks.}
   \label{FIG:tJ}
\end{figure}

Figure~\ref{FIG:structure} shows the schematics of various lattice configurations with real-space unit cells used in this work for \FeSe.
Recognizing the crystallographic similarities between the isostructural FeSe-122 and FeAs-122 compounds and that
the electron pocket at the $M$ point is not only a common but also main feature in the heavily electron-doped region, $0\le y \le 0.4$,
we use the same tight-binding model for the electron hopping as in Refs.~\onlinecite{DZhang,TZhou}.
We account for the reported electronic band structure of the perfect lattice of K$_x$Fe$_2$Se$_2$,\cite{YZhang01}  shown in Fig.~\ref{FIG:tJ}(a),
by proposing a modified set of hopping parameters $(t_1, t_2, t_3, t_4) =(1, 1, -2, 0.08)$.

In our approach, we use for simplicity the same hopping parameters for K$_{0.8}$Fe$_{1.6}$Se$_2$ as for all other K$_x$Fe$_2$Se$_2$ compounds.
The unit cell of K$_{0.8}$Fe$_{1.6}$Se$_2$ is modified to a $\sqrt{10}\times\sqrt{10}$ area, see Fig.~\ref{FIG:structure}(a) with cut  ``-c-'' for the unit cell,
due to the periodic vacancy order along stripes at the doping concentration $y=0.4$.
In this real-space unit cell there are a total of 10 sites,  namely, 8 iron atoms and 2 vacancies.
We start with an effective lattice model for K$_{0.8}$Fe$_{1.6}$Se$_2$ by including the hopping $H^t$ and exchange interaction $H^{J}$ terms:
\begin{equation}
	H = H^t + H^{J},
\end{equation}
where 
\begin{equation}
	H^t = -\sum_{i j \mu\nu\sigma} t_{i \mu j \nu} c^\dagger_{i\mu,\sigma}c_{j\nu,\sigma}
	- t_0 \sum_{i \mu \sigma}c^\dagger_{i\mu,\sigma}c_{i\mu,\sigma} ,
\end{equation}
and
\begin{equation}
\begin{aligned}
	H^{J} &= J_1\sum_{<ij>\mu\nu} {\vec S}_{i\mu}\cdot{\vec S}_{j\nu} + J_2\sum_{\ll ij\gg\mu\nu} {\vec S}_{i\mu}\cdot{\vec S}_{j\nu} \\
			&+ J_1'\sum_{<ij>'\mu\nu} {\vec S}_{i\mu}\cdot{\vec S}_{j\nu} + J_2'\sum_{\ll ij\gg'\mu\nu} {\vec S}_{i\mu}\cdot{\vec S}_{j\nu} .\\
\end{aligned}
\end{equation}
The hopping parameters on the lattice are defined as
\begin{equation}
	\begin{tabular}{ c| c | c | c}
	  $t_{<ij>,\mu=\nu}$ & $t_{\ll ij\gg,\mu=\nu}^{upper}$ & $t_{\ll ij\gg,\mu=\nu}^{lower}$  & $t_{\ll ij\gg,\mu\neq\nu}$ \\ \hline
	  $t_1=1$ & $t_2=1$ & $t_3=-2$ & $t_4=0.08$\\
	\end{tabular}
\end{equation}
where $i\, (j)$ are site indices, and $\mu \,(\nu)$ are orbital indices corresponding to $d_{xy} \mbox{ or } d_{yz}$ wave function orbitals and
$t_0$ is the chemical potential.
The expressions $<ij>$ ($\ll ij\gg$) and $<ij>'$ ($\ll ij\gg'$) denote intra- (inter)-block nearest-neighbor ($nn$) and next-nearest-neighbor ($nnn$) 
hopping processes, whereas
$t^{upper}$ ($t^{lower}$) indicates the hopping term crossing over the upper (lower) Se atom as shown in Fig.~\ref{FIG:tJ}(a).
If we further approximate the exchange interaction to be of the Ising type then only the $S^z$ component is involved 
\begin{equation}
\begin{aligned}
	S_{i\mu}^z 	&= \frac{1}{2}\sum_{\alpha\alpha'} c^\dagger_\alpha \sigma^z_{\alpha\alpha'} c_{\alpha'},
\end{aligned}
\end{equation}
and the interaction term can be expressed in mean-field approximation by
\begin{equation}
\begin{aligned}
	S_{i\mu}^z S_{j\nu}^z	&= \frac{1}{4}(\langle n_{i\mu\uparrow}\rangle-\langle n_{i\mu\downarrow}\rangle)(n_{j\nu\uparrow}-n_{j\nu\downarrow})\\
							&+ \frac{1}{4}(\langle n_{j\nu\uparrow}\rangle-\langle n_{j\nu\downarrow}\rangle)(n_{i\nu\uparrow}-n_{i\nu\downarrow}) .
\end{aligned}
\end{equation}
In Fig.~\ref{FIG:tJ}(b) the $nn$ intra- (inter)-block exchange term $J_1$ ($J_1'$) and $nnn$ intra- (inter)-block exchange term $J_2$ ($J_2'$) are illustrated.
We can now construct the corresponding mean-field Bogoliubov-de Gennes (BdG) matrix equation on a lattice:
\begin{equation}
	\sum_{j\nu} 
		\left( \begin{array}{cc}
		H_{i\mu j\nu\uparrow}	& \Delta_{i\mu j\nu} \\
		\Delta_{i\mu j\nu}^*	&-H_{i\mu j\nu\downarrow}
		\end{array} \right)\
		\left( \begin{array}{c}
		u^n_{j\nu\uparrow} \\
		v^n_{j\nu\downarrow} 
		\end{array} \right)\ = E_n
		\left( \begin{array}{c}
		u^n_{i\mu\uparrow} \\
		v^n_{i\mu\downarrow} 
		\end{array} \right)\ .
		\label{EQ:BdG}
\end{equation}
Here the single-particle Hamiltonian $H_{i\mu j\nu,\sigma}$ is expressed by
\begin{equation}
\begin{aligned}
	H_{i\mu j\nu\sigma} = -t_{i\mu j\nu}+\frac{\sigma}{4}\sum_{\delta\mu}[(J_{\delta} +J_{\delta}')\langle m_{i+\delta,\mu} \rangle ]\delta_{ij}\delta_{\mu\nu}\\
-t_0\delta_{ij}\delta_{\mu\nu} ,
\end{aligned}
\end{equation}
where $\sigma$ correspond to $\pm 1$ for spin-up(down) index, $\delta$ is taken with $\pm \hat x$ ($\pm \hat y$) and $\pm \hat x \pm \hat y$, 
which correspond to the $nn$ and $nnn$ real space shift; 
$J$  ($J'$) is for the intra (inter) block interaction and the mean-field magnetization per site and per orbital is
$\langle m_{i\mu}\rangle = \mu_B( \langle n_{i\mu\uparrow}\rangle-\langle n_{i\mu\downarrow}\rangle)$.
The quasiparticle energies $E_n$ are measured with respect to the chemical potential.  
We note that the single-particle Hamiltonian depends on spin- and orbital-dependent electron density and pairings, which are given by
\begin{eqnarray}\label{spin_occupations}
		\langle n_{i\mu\uparrow}	\rangle &=& \sum_{n} |u_{i\mu\uparrow}^n|^2 f(E_n)\;,
	\\
		\langle n_{i\mu\downarrow}	\rangle &=& \sum_{n} |v_{i\mu\downarrow}^n|^2 [1-f(E_n)]\;,
	\\
		\Delta_{i\mu j\nu} &=& \frac{V_{i\mu j\nu}}{4} \sum_n (u^n_{i\mu\uparrow}v^{*n}_{j\nu\downarrow}+u^n_{j\nu\uparrow}v^{*n}_{i\mu\downarrow})\tanh \frac{E_n}{2 T} .
		\label{gap}
\end{eqnarray}
Here we set the Boltzmann constant $k_B=1$. 
Finally, the BdG equation~(\ref{EQ:BdG}) must be solved self-consistently with Eqns.~(\ref{spin_occupations})-(\ref{gap}).
Since we consider only normal-state properties in this work,  we ignore the superconducting pairing term, $V_{i\mu j\nu}=0$, in the BdG matrix equation.

\section{Electronic structure in the paramagnetic state}
\label{SEC:Bloch}

\begin{figure}
 \includegraphics[scale=0.55,angle=0]{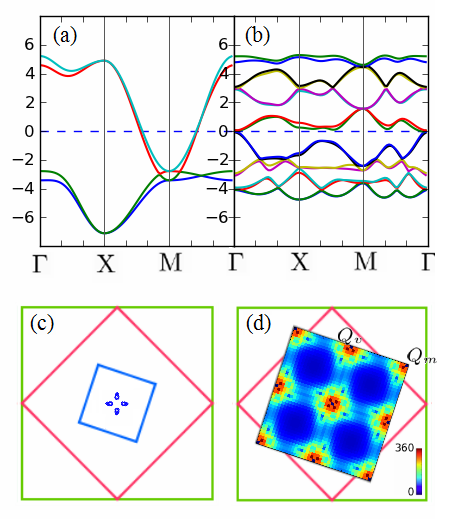}
  \caption
   {(Color online) The electronic band structure for compounds
   (a) K$_{0.8}$Fe$_2$Se$_2$ (fill factor $n=2.4$, $t_0=1.09$ for high electron doping) and
   (b) K$_{0.8}$Fe$_{1.6}$Se$_2$ (half-filling $n=2.0$, $t_0=0.005$) with $(t_1, t_2, t_3, t_4) = (1, 1, -2, 0.08)$.
   The dashed blue line is the location of the chemical potential.
   (c) Fermi surface of electron pockets of K$_{0.8}$Fe$_{1.6}$Se$_2$ located near the $\Gamma$ point in $k$-space corresponding to the $\sqrt{10}\times\sqrt{10}$ unit cell (blue  square).a
   (d) Static spin susceptibility for the bare band structure of K$_{0.8}$Fe$_{1.6}$Se$_2$.
      The high intensity spots at scattering wave vectors $Q_v=\pi(\frac{1}{5},\frac{3}{5})$ and $Q_m=\pi(\frac{4}{5},\frac{2}{5})$ are in agreement with neutron scattering experiments.} 
   \label{FIG:band}
\end{figure}

The most common way to construct the electronic band structure of ordered systems is to use the Bloch wave function formulation by
In case of the enlarged $\sqrt{10}\times\sqrt{10}$ unit cell with 8 Fe atoms and 2 vacancies,
the corresponding {\bf k}-space Hamiltonian is a $16\times16$ matrix, $H^t_{16\times 16}$, which is straightforward but tedious to derive.
Instead, we propose another method based on the impurity problem solution that produces exactly the same results and, moreover, provides physical
insight into the breaking of the translational symmetry and how it gradually affects the electronic structure with increased vacancy doping.
In this approach the vacancy is mapped onto an impurity with an adjustable onsite scattering potential $V_0$ that varies from
$0$ (no vacancy) to $\infty$ (strong scattering center).
Also the continuous tunability of the impurity potential provides an easy handle on the evolution of the electronic structure with scattering strength.
Please refer to the Appendix for more details on the implementation of the impurity problem calculation employed in this work.

In Fig.~\ref{FIG:band}(a) we show the electronic band structure of K$_{0.8}$Fe$_2$Se$_2$ for our tight-binding model parameterization
using the $\sqrt{2}\times\sqrt{2}$ unit cell (cut ``-b-'' in Fig.~\ref{FIG:structure}(a)). The corresponding band structure of the vacancy ordered compound K$_{0.8}$Fe$_{1.6}$Se$_2$  is shown for comparison in Fig.~\ref{FIG:band}(b).
The evolution of the dispersion from the $\sqrt{2}\times\sqrt{2}$ unit cell with four bands to the enlarged $\sqrt{10}\times\sqrt{10}$ unit cell with 16 bands is nontrivial and cannot be obtained by a simple rigid-band shift of the chemical potential. Indeed,  the Fermi surface topology for K$_{0.8}$Fe$_{1.6}$Se$_2$  shows very tiny electron pockets located near the $\Gamma$ point, see Fig.~\ref{FIG:band}(c), in contrast to K$_{0.8}$Fe$_2$Se$_2$.

In addition to the electronic dispersion, we calculate the static spin susceptibility $\chi({\bf q})$ for the bare band structure of itinerant electrons,
\begin{eqnarray}
 \chi({\bf q})=\sum_{IJ}\chi_{IJ}({\bf q},i\Omega_m \rightarrow 0)\;,\\
 \chi_{IJ} = \frac{1}{N}\langle T_\tau [S^z_I(\mathbf{q},\tau)S^z_J(-\mathbf{q},0)]\rangle,
\end{eqnarray}
where $N$ is the total number of unit cells and  $S^z_I(\mathbf{q},\tau)$ is the Fourier component corresponding to the site and orbital index $I$ in the unit cell.
In this compact notation, the orbital wave functions have the running super-indices $I=(i,\mu)$ and $J=(j,\nu)$. 
The dynamic susceptibility is given by
\begin{eqnarray}
  \chi_{IJ}({\bf q},i\Omega_{m})& =& - \frac{T}{4N}\sum_{\mathbf{K}\alpha}\sum_{n}G_{J\alpha,I\alpha}(\mathbf{K},i\omega_{n}) \nonumber \\
   && \times
 G_{I\alpha,J\alpha}({\bf K}+{\bf q},i\omega_{n}+i\Omega_m)\; ,
 \label{EQ:susceptibility}
\end{eqnarray}
where the wave vector $\mathbf{K}$ is defined in the Brillouin zone corresponding to the 
$\sqrt{10}\times\sqrt{10}$ unit cell, $\omega_{n}=(2n+1)\pi T$ and $\Omega_m = 2m \pi T$ are the Matsubara frequencies of the fermions and bosons. In standard notation the multiorbital lattice Green's functions are given by 
\begin{subequations} 
\begin{equation}
G_{I\uparrow,J\uparrow}(\mathbf{K},i\omega_{n})= \sum_{n} \frac{u_{I\uparrow}^{n}({\bf K})\; u_{J\uparrow}^{n*}({\bf K})}{i\omega_{n}-E_n(\bf K)}\;,
\end{equation}
\begin{equation}
G_{I\downarrow,J\downarrow}(\mathbf{K},i\omega_{n})	=\sum_{n}  \frac{v_{I\downarrow}^{*n}({\bf K})\; v_{J\downarrow}^n({\bf K})}{i\omega_{n}+E_n(\bf K)}\;.
\end{equation}
\label{EQ:Green}
\end{subequations}
In the calculation of the itinerant spin susceptibility the red spots in Fig.~\ref{FIG:band}(d) show high intensity  around  ${\bf q} = Q_\star$ and $Q_\triangle$
in agreement with neutron scattering experiments.\cite{WBao:2011}
It follows from the Stoner criterion that the observed AFM state is possibly formed from itinerant electrons of the paramagnetic state due to the bare band structure, rather than the exchange interaction of localized spins. Hence it is natural to expect that close to the magnetic instability a small driving force can break the symmetry of the paramagnetic state and induce the 
long-range AFM state.

\section{Electronic structure of random vacancy lattice}
\label{SEC:random}

\begin{figure}  
 \includegraphics[width=85mm, angle=0]{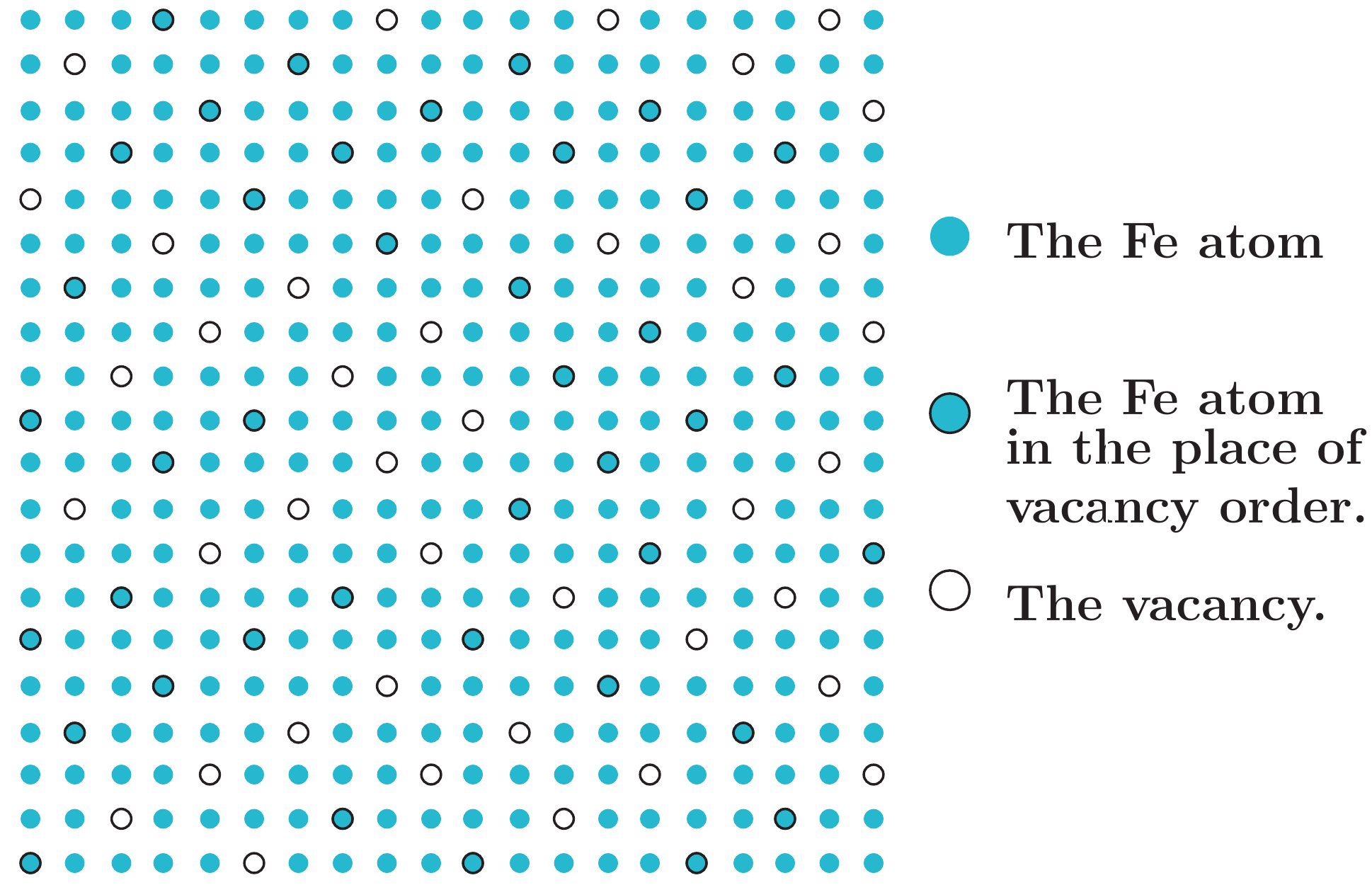}
  \caption
   {
   (Color online) Schematic picture of a $20\times20$ lattice with random vacancies.
   For this configuration the minimum size of a supercell is a $10\times 10$ square lattice.
   This supercell contains the translational symmetries of both the $\sqrt{2}\times\sqrt{2}$ and $\sqrt{5}\times\sqrt{5}$ unit cells.
   The random lattice is constructed by randomly removing vacancies from the ordered diagonal stripes of phase-separated vacancies.
   The phase separation pattern along the direction (2,1,0) becomes clearer for larger superlattices.
   }
   \label{FIG:random}
\end{figure}

The periodic Bloch wave function formulation becomes questionable for lattices with random vacancy disorder in the intermediate doping region $0 < y < 0.4$,
where a perturbative approach to disorder in supercells fails.
Here, we describe the construction of a random vacancy lattice calculation (RLC).
First, we construct a $40\times40$ or larger real-space lattice for A$_x$Fe$_2$Se$_2$ with vacancies phase-separated in stripes along the (2,1,0) direction as shown in Fig.~\ref{FIG:random}.
Second, we randomly add iron atoms, i.e., remove vacancies, on the vacancy ordered stripes up to the point that we match the vacancy occupation for compounds measured by ARPES.
Third, we construct the corresponding real-space Hamiltonian $H^t$ with hopping terms $(t_1, t_2, t_3, t_4) = (1, 1, -2, 0.08)$.
Finally, we exactly diagonalize the Hamiltonian and compute its pairs of eigenvalues $E_n$ and eigenvectors $(u_{i\mu \uparrow}^{n},  v_{i\mu\downarrow}^n)$ from which
we calculate the lattice Green's functions. 

When going beyond the dilute limit of random disorder, as for example in a heavily doped alloy, the randomness of vacancies plays a crucial role for the electronic dispersion, because
the rigid-band shift appropriate in the dilute limit  breaks down.
Hence the conventional approach of periodicity and the calculation of the electronic band structure and Fermi surface fail.
In order to solve this complicated problem, we calculate the multiorbital real-space spectral function 
$A_{I J}(\omega)=-\frac{1}{\pi}{\rm Im}\, G_{I J}(i\omega_n\rightarrow \omega + i\Gamma)$, 
which depends on the multiorbital real-space Green's function summed over all bands $n$,
\begin{equation}
  G_{IJ}(i\omega_n) =  G_{I\uparrow,J\uparrow}(\mathbf{K}=0,i\omega_n) + G_{I\downarrow,J\downarrow}(\mathbf{K}=0,i\omega_n)\;,
\end{equation}
where the general form of the spin-dependent Green's functions on the
right-hand side of the above equation has been given  in Eq.~(\ref{EQ:Green}).
Finally, the Fourier transform of $A_{I J}(\omega)$ to {\bf k}-space for the one-iron per unit cell (1-Fe) gives the desired spectral function, which is the one
measured in ARPES experiments,
\begin{equation}
  A({\bf k}, \omega) = \sum_{I J} A_{IJ}(\omega) \exp\{-i{\bf k}\cdot({\bf R}_i-{\bf R}_j)\} .
\end{equation}
Here, i (j) are lattice indices for the location of each Fe atom in the $40\times40$ supercell with orbital indices
$\mu$ ($\nu$) and super-indices $I=(i,\mu)$ and $J=(j,\nu)$.
The double-sum is a short-hand notation for $\sum_{IJ} \equiv \sum_{i j}\sum_{\mu \nu}\delta_{\mu \nu}$. Since we considered only the bare band structure, we dropped for convenience the spin indices in this calculation.

\begin{figure}  
 \includegraphics[width=85mm,angle=0]{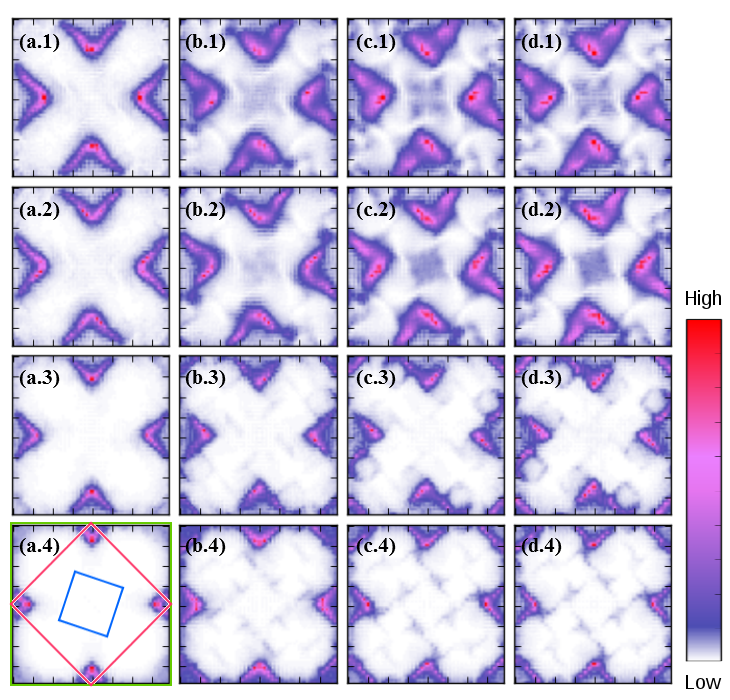}
  \caption
   {(Color online) The calculated Fermi surface evolution from the Fourier transformed real-space spectral function
   in the 1-Fe BZ. The alphabetical order in each picture represents different vacancy concentrations:
   (a.$n$) Fe$_{1.9}$, (b.$n$) Fe$_{1.78}$, (c.$n$) Fe$_{1.72}$, (d.$n$) Fe$_{1.7}$, while the
   numerical index $n=1-4$ represents different values of the chemical potential.
   The chemical potential varies from high to low as $n$ increases: 
   ($t_0^{(1)},t_0^{(2)},t_0^{(3)},t_0^{(4)})=(0.5,0.33,-0.7,-1.5)$.
   The calculations in panels (b.1), (c.2) and (d.3) correspond to the compounds
   (Tl,K)Fe$_{1.78}$Se$_2$, Tl$_{0.58}$Rb$_{0.42}$Fe$_{1.72}$Se$_2$ and K$_{0.8}$Fe$_{1.7}$Se$_2$, respectively.
   In panel (a.4) we redraw the different unit cells of Fig.~\ref{FIG:structure}(b) with
   $1\times1$ (green), 
   $\sqrt{2}\times\sqrt{2}$ (red),  
   and $\sqrt{10}\times\sqrt{10}$ (blue). 
   }
   \label{FIG:FS}
\end{figure}

Next we apply the RLC method to calculate for both doping variables $x$ and $y$ the evolution of the spectral function and Fermi surface topology for compounds \FeSe\ . For the hopping term, H$^t$, the variables $x$ and $y$ can be mapped onto the value of the chemical potential t$_0(x)$and vacancy concentration $y$. 
Figure~\ref{FIG:FS} shows the calculated Fermi surface evolution for changing t$_0(x)$ and $y$ values on random vacancy disorder lattices. These lattices were already large enough to self-average the vacancy disorder, thus no further ensemble average of different random vacancy configurations was required.
The visualization of the evolution of the spectral function helps us to understand when an electron pocket ``appears'' at the $\Gamma$ point, namely for $y\ge 0.22$, see panels (b)-(d), as well as when it disappears for shifted chemical potentials $t_0^{(3)} = -0.7$ and $t_0^{(4)}=-1.5$ (rows 3 and 4). 
Our finding differs from that discussed in Ref.~\onlinecite{WeiLi:2012}, where it was claimed that the electron pocket at the $\Gamma$ point of the FeSe-122 compound originates from the BAFM  structure. However, our results clearly demonstrate the presence of an electron pocket at $\Gamma$ in the absence of BAFM order and over a large range of vacancy concentrations $0.22 \le y \le 0.4$.
Indeed  Fig.~\ref{FIG:FS} shows that the small electron pocket at the $\Gamma$ point originates from the same regions in the Brillouin zone as for the vacancy stripe-ordered compound with Fe$_{1.6}$ shown in Fig.~\ref{FIG:band}.
These calculations demonstrate the power of the RLC method, when randomness (doping) strongly affects the Fermi surface. This computational method is adequate to understand the evolution of the electronic structure for intermediate doping levels in the range $0 \le y \le 0.4$, where the stable electronic structure is a compromise between the solution for the perfect lattice with $y=0$ and the stripe-ordered lattice with $y=0.4$. 
Alternatively one can obtain these results by diagonalizing large supercells with random lattice Hamiltonians. Of course such a brute force approach is time consuming, especially when many calculations for different random configurations are required as illustrated in the cartoon of Fig.~\ref{FIG:random}.
We confirmed numerically our results by sampling many random vacancy configurations of large supercells and found that the spectral functions are no different from the RLC method presented here for fixed iron concentrations.
In Table~\ref{table2} we list results for both periodic lattice calculations (PLC) and random vacancy lattice calculations (RLC) performed for various vacancy concentrations $y$.
\newcommand{\TlRb}	{ Tl$_{0.58}$Rb$_{0.42}$Fe$_{1.72}$Se$_{2}$}
\newcommand{\TlK}	{ (Tl,K)Fe$_{1.78}$Se$_{2}$}
\newcommand{\AFS}	{ A$_{x}$Fe$_{2-y}$Se$_{2}$}
\newcommand{\KFS}	[3]{ K$_{#1}$Fe$_{#2}$Se$_{#3}$}
\begin{table}
\caption{The FeSe-122 compounds and their respective vacancy doping $y$, whose Fermi surface topology was fitted to local density approximation (LDA) calculations or ARPES measurements.}\label{table2}
\begin{center}
	\begin{tabular}{ l| r| r| r | c}
	\hline\hline
	  ${y}$		& compound			& fitting	& lattice	& electron pockets\\ \hline
	  0.00		&\KFS{0.8}{2}{2}	& LDA		& PLC		& M\\
	  0.22		&\TlK				& ARPES		& RLC		& M\\	
	  0.28		&\TlRb				& ARPES		& RLC		& M \& $\Gamma$\\
	  0.30		&\KFS{0.8}{1.7}{2}	& ARPES		& RLC		& M\\
	  0.40		&\KFS{0.8}{1.6}{2}	& LDA		& PLC		& $\Gamma$\\ \hline\hline
	\end{tabular}
\end{center}
\end{table}

\section{Magnetic structure in the AFM state}
\label{SEC:AFM}

\begin{figure}  
 \includegraphics[scale=0.36,angle=0]{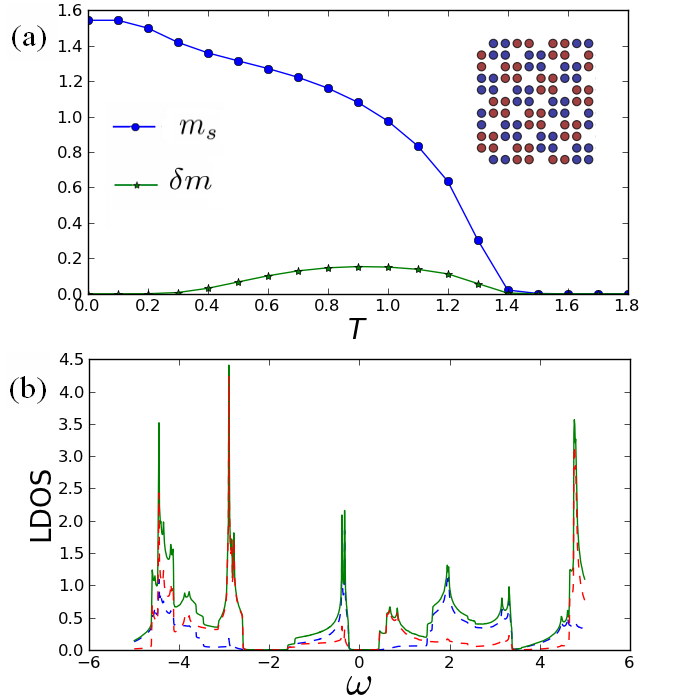}
  \caption{(Color online) (a) Temperature dependence of the magnetization (as defined in the main text)  in the BAFM state. The temperature $T$ is in units of $t_1$.
  The inset shows a small part of the lattice used in  our mean-field calculation ($20\times20$ sites with 20\% vacancies).
(b) The zero-temperature LDOS as a function of energy $\omega$ in units of $t_1$. Shown are the
  two block sublattices  with net spin up (dashed red) and down (dashed blue) lines as well as their sum (solid green). 
A gap of $\sim 0.4\, t_1$ opens  at the Fermi level $\omega=0$. }
\label{FIG:BAFM}
\end{figure}

Now that we have fully developed the parameterization for the kinetic part $H^t$ of the $t$-$J$ lattice model for random vacancy order with a single set of hopping parameters, we can focus on the exchange interaction term $H^J$.
We start with a mean-field calculation for the magnetic state, using a set of rescaled exchange parameters extracted from the band structure calculations.~\cite{CCao:2011b} 
We rescale the {\it ab}-{\it initio} exchange parameters by a global factor to obtain $(J_1, J_2, J_1', J_2')= (-3.44,  -0.36,  -1.16,  1.52)t_1$, which results in better agreement of the calculated N\'eel temperature with experiment.  Within our low-energy effective $t$-$J$ model, this set of exchange parameters  leads to a N\'eel temperature for the block-spin AFM state, $T_N\approx 1.4 \, t_1 \approx -0.407\, J_1$, see Fig.~\ref{FIG:BAFM}(a).  If we assume a hopping integral $t_1=35$ meV, we obtain $T_N=569$ K. 
Indeed this value is very close to the one observed experimentally ($T_N=559$ K).\cite{WBao:2011} 
We argue that this quantitative result provides further evidence for the dominance of the kinetic term over the exchange term in the $t$-$J$ Hamiltonian with incipient AFM order due to Fermi surface nesting.
Furthermore, we calculate the staggered magnetization as defined by 
\begin{eqnarray}
m_{s} &=& \frac{1}{N_{L}}\sum_{i} \vert m_i \vert\;, 
\end{eqnarray}
and its deviation from the standard AFM state may be defined as
\begin{eqnarray}
\delta m &=& \frac{2}{N_{L}}\vert \sum_{i}  m_i \vert\;. 
\end{eqnarray}
Here $N_L$ is the total number of Fe sites and 
$m_i = \sum_{\mu} \langle m_{i\mu}\rangle$ 
is the spin density on each site $i$, 
where the summation index $i$ runs over all lattice sites. In the BAFM state, the system breaks the translational symmetry in the $\sqrt{5}\times\sqrt{5}$ unit cell and gives rise to a larger periodicity with a $\sqrt{10}\times\sqrt{10}$ unit cell.
The result for the magnetization as a function of temperature is shown in Fig.~\ref{FIG:BAFM}(a),
which is in qualitative agreement with the neutron scattering experiments,
except for a smaller magnetic moment as observed experimentally per iron site ($3.3 \, \mu_B$).~\cite{WBao:2011}
The reason for this discrepancy follows from our phenomenological two-orbital model, which has only 2 electrons per Fe site. Therefore, the maximum total moment on each site is 2$\mu_B$. Whereas all 5 electrons of the Fe atom seem to participate in the moment formation.
Notably our result shows an antiferrimagnetic state at finite temperatures for $T>0.3 t_1$, where $\delta m \neq 0$, that is, the magnetization for the up and down spin blocks are slightly different leading to a net ferri magnetization.
This difference is related to the fact that in the present model an asymmetry for the intra-orbital next-nearest-neighbor hopping integrals $t_2$ vs. $t_3$ has been introduced due to the selenium atoms below and above the iron layer. We confirmed numerically that for $|t_2|=|t_3|$ a purely antiferromagnetic state exists. It is interesting to note that this type of difference in kinetic hopping terms also manifests itself in the magnetization between the two spin blocks of the bipartite sublattices.
The existence of antiferrimagnetism could be identified by  weak satellite peaks at the ferromagnetic propagating wave vector $\mathbf{Q}=0$ in elastic neutron scattering of K$_{0.8}$Fe$_{1.6}$Se$_2$ for temperatures  $0.3 t_1 <T<T_N$. Therefore using the same arguments one should also expect an asymmetry in the AFM split satellite peaks of the nuclear magnetic resonance (NMR) spectrum.
The observation of antiferrimagnetism could serve as an important evidence for different Fe 3$d$-electron hopping integrals mediated by Se atoms buckling on both sides of the Fe layer in the bulk system.

Finally, we calculate the local density of states (LDOS)  given by
\begin{equation}
	\rho_i(\omega) = \sum_{n\mu} [|u^n_{i\mu\uparrow}|^2\delta(E_n-\omega)+|v^n_{i\mu\downarrow}|^2\delta(E_n+\omega)],
\end{equation}
where the delta-function $\delta(x)$ is approximated as $\delta(x)\approx \Gamma/\pi(x^2+\Gamma^2)$ with the intrinsic quasiparticle broadening parameter $\Gamma$. In our calculation, we  choose  $\Gamma=0.002$  but note that the result does not change qualitatively for small changes of $\Gamma$. 
For the LDOS calculations we use an $80\times80$ sites supercell. 
Figure~\ref{FIG:BAFM}(b) shows the zero-temperature LDOS in the BAFM phase on two 
block sublattices. We note that the LDOS on each site of a chosen block is the same. As a consequence of the exchange interaction a clear gap of size $\sim0.4 t_1$  
opens at the Fermi energy $\omega=0$, which is in qualitative agreement with {\it ab}-{\it initio} band structure calculations for the BAFM state.~\cite{CCao:2011b}

\section{Conclusion}
\label{SEC:Conclusion}

In summary, we systematically addressed within an effective two-orbital model the effects of the iron vacancy on the normal-state electronic structure and magnetic properties in 
the A$_x$Fe$_{2-y}$Se$_2$ compounds.  We determined a suitable choice of hopping parameters for Fe $3d$-electrons by fitting the Fermi surface topology of ARPES data.   From the determined hopping parameters we calculated the evolution of the electronic structure and mapped out the Fermi surface topology for arbitrary Fe vacancy concentration based on a random vacancy disorder model.  After the kinetic part of the $t$-$J$ Hamiltonian was parameterized we focused on the special case of A$_{0.8}$Fe$_{1.6}$Se$_2$, where the Fe atoms form the $\sqrt{5}\times \sqrt{5}$ vacancy order. Here we studied the magnetic properties  by including the exchange interactions with relative strengths extracted from {\it ab}-{\it initio} calculations. Our mean-field solution of the $t$-$J$ Hamiltonian reproduced the block-spin antiferromagnetic (BAFM) state, in good agreement with neutron scattering experiments and {\it ab}-{\it initio} calculations. Finally, we found that the magnitude of the magnetization on the two-block bipartite sublattices is at a small variance for temperatures $0.3 t_1 < T < T_N$. This feature is unique to our model with different Se-mediated hopping strengths (below and above the Fe layer) on the two block spin sublattices, which can be tested by future refined neutron scattering and nuclear magnetic resonance experiments in the corresponding temperature range.

\begin{acknowledgments}
We thank Alexander Balatsky, Tanmoy Das, Yi Gao, Jian Li, and Wei Li for helpful discussions.
One of us (Y.-Y.T.) acknowledges the hospitality of the Los Alamos National Laboratory, where part of this work was carried out.
This work was supported by the Robert A. Welch Foundation under Grant. No. E-1146 and the Los Alamos National Laboratory through  the Basic Energy Sciences program and the UC Laboratory Fees Research program under the U.S.\ DOE contract no.~DE-AC52-06NA25396. The results  were first reported in Ref.~\onlinecite{yTai:2012}.
\end{acknowledgments}

\appendix
\section{Treatment of the vacancy impurity problem}
\label{Appendix}

\begin{figure}  
 \includegraphics[scale=0.28,angle=0]{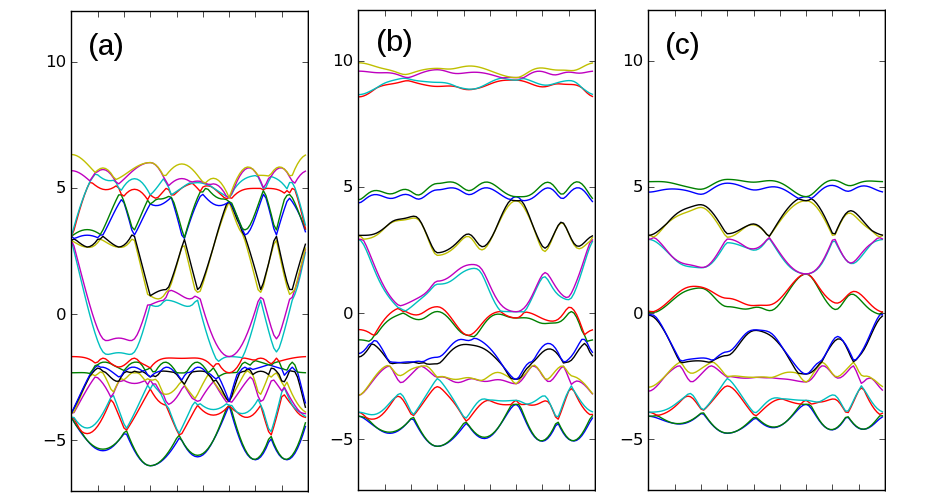}
  \caption
   {(Color online) The band structure plot in high symmetry axis ($\Gamma\rightarrow X\rightarrow M\rightarrow\Gamma$)
   for  $V_0=0$ (a),  $V_0=2.5$ (b),  $V_0=100$ (c).}
   \label{FIG:impurityModel}
\end{figure}

We treat the vacancy disorder as an impurity scattering problem, since one might anticipate that a vacancy behaves like a strong impurity scatterer for a Bloch wave function.
Therefore, we put an on-site impurity energy $\epsilon_i$  on each vacancy site $i=i_{v}$
to mimic vacancy disorder:
\begin{equation}
\begin{aligned}
  H	&= H_{hop} + H_{imp} \\
 	 &=\sum_{i,j} (-t_{ij}-\mu\;\delta_{ij})\;c^\dagger_{i} c_{j}+\sum_{i=i_{v}}V_0 \;c^\dagger_{i} c_{i},
\end{aligned}
\end{equation}
We introduce a new index to describe the $A_{1(2)};B_{1(2)}$ sub-atoms for unit cell cut ``-b-'' in Fig.~\ref{FIG:structure}, $\psi^\dagger_i = (c^\dagger_{A1},c^\dagger_{A2},c^\dagger_{B1},c^\dagger_{B2})_i$,
and linearize the fermionic operator, $c_{A\alpha(B\alpha),i}=(1/\sqrt{N})\sum_{\bf k} c_{A\alpha(B\alpha),\bf k} \times \exp(i{\bf k}\cdot R_i)$, to get the $4\times 4$
hopping matrix in {\bf k}-space, $H_{hop} = \sum_{\bf k} \psi^\dagger_{\bf k} M_{\bf k} \psi_{\bf k}$,
\begin{equation}
  M_{\bf k} = 
  \left(
	\begin{array}{ccccc}
	 \varepsilon_{A,\bf k}-\mu	& \varepsilon_{xy,\bf k} & \varepsilon_{T,\bf k} & 0 \\
	 \varepsilon_{xy,\bf k}	& \varepsilon_{A,\bf k}-\mu  & 0				 & \varepsilon_{T,\bf k} \\
	 \varepsilon_{T,\bf k}	& 0					 & \varepsilon_{B,\bf k}-\mu & \varepsilon_{xy,\bf k}\\
	 0					& \varepsilon_{T,\bf k}	 & \varepsilon_{xy,\bf k}& \varepsilon_{B,\bf k}-\mu
	\end{array}
  \right)
\end{equation}
We perform the same linear transformation for the impurity term $H_{imp}$ in the Hamiltonian. There are four independent vacancy ordering vectors as 
shown in the unit cell cut ``-b-''  in Fig.~\ref{FIG:structure}
for $Q_1=2\pi(4/5,2/5), Q_n = nQ_1, n=1 \cdots 4$.\cite{Das2011}

By construction $H_{imp}$ enlarges the basis in the {\bf k}-space, $\psi_{\bf k} \rightarrow \psi_{\bf k}'=(\psi_{\bf k},\psi_{\bf k+Q_1},\psi_{\bf k+Q_2},\psi_{\bf k+Q_3},\psi_{\bf k+Q_4})$.
In this formulation the diagonal scattering matrix $V=I_{4\times 4}\times V_0$  sits in the off-diagonal positions to describe multiple scattering between all vacancy ordered states with different $Q_n$,   Finally, the  enlarged Hamiltonian becomes $H = \sum_{\bf k} \psi'^\dagger_{\bf k} W_{\bf k} \psi'_{\bf k}$ with
\begin{equation}
W_k = 
\left(
\begin{array}{ccccc}
 M_{\bf k} & V & V & V & V \\
 V & M_{\bf k+Q_1} & V & V & V \\
 V & V & M_{\bf k+Q_2} & V & V \\
 V & V & V & M_{\bf k+Q_3}  & V \\
 V & V  & V & V & M_{\bf k+Q_4}
\end{array}
\right)
\end{equation}
Note that the impurity potential $V_0$ not only leads to a reconstruction of the shape of the band structure,
it also creates a gap of $2V_0$ between high-energy bands and the low-energy bands.
Therefore, we need to shift it back to the origional place, $W_k \rightarrow W'_k=W_k+I_{20\times 20}\times V_0$.\\\indent

Figure~\ref{FIG:impurityModel} shows the calculated band structure for various  values of the impurity scattering strength $V_0$. When $V_0=0$, the band structure exhibits an entanglement of 20 bands, which down-folded simplifies to the band structure shown for the four bands in the $\sqrt{2}\times\sqrt{2}$ unit cell of
Fig.~\ref{FIG:structure}(a).
The more complicated plot in 
Fig.~\ref{FIG:impurityModel}(a) it is due to the repeated plotting of different $Q_{n}$'s in {\bf k}-space of $M_{\bf k}$.
However, for finite values of $V_0$ a gap separates the upper four upper bands and 16 lower bands, shown in Fig.~\ref{FIG:impurityModel}(b).
Finally, when $V_0$ approaches the unitarity limit. $V_0 \to \infty$, as shown in Fig.~\ref{FIG:impurityModel}(c), the upper four bands are being pushed far above
as four independent flat bands, while the lower 16 bands form a simpler shape as the new periodicity with vacancy stripe order sets in.

\bibliography{AFe2Se2}

\end{document}